\documentclass[a4paper,11pt]{article}
\pdfoutput=1 

\usepackage{jheppub}

\usepackage[mmddyyyy,hhmmss]{datetime}

\usepackage{amsmath,graphicx,amssymb,subfigure,bbm,comment,mathtools}
\usepackage[all]{xy}
\usepackage{slashed}

\begin{document}


\title{Giant Gravitons on the Schr\"{o}dinger pp-wave geometry}

\author[a]{George Georgiou}
\author[a,b]{and Dimitrios Zoakos}

\affiliation[a]{Department of Physics, National and Kapodistrian University of Athens, 15784 Athens, Greece}
\affiliation[b]{Hellenic American University, 436 Amherst st, Nashua, NH 03063 USA}

\emailAdd{ggeo@phys.uoa.gr}
\emailAdd{zoakos@gmail.com}

\abstract{We construct a new giant graviton solution on the recently constructed pp-wave geometry of 
the non-supersymmetric Schr\"{o}dinger background. That solution exhibits an intriguing behavior as the deformation 
parameter of the spacetime varies. Firstly, the degeneracy between the giant and the point graviton is lifted 
for the benefit of the giant graviton as soon as the deformation is turned on. Secondly, when the deformation 
parameter exceeds a critical value the barrier separating the point from the giant graviton disappears. This suggests that the mere presence of a D3-brane leads to the spontaneous breaking of 
conformal invariance. We perform a detailed analysis of the full bosonic spectrum, which reveals that the deformation 
induces a coupling between the scalar and the gauge field fluctuations. It is exactly this coupling that keeps the 
giant graviton free of tachyonic instabilities. Furthermore, the giant graviton configuration 
completely breaks the supersymmetry of the pp-wave background, as the Kappa-symmetry analysis suggests.}


\maketitle
\flushbottom


\section{Introduction}

Finding the spectrum of strings propagating on a generic curved background is a notoriously difficult task. At the same 
time, it is an extremely important task because of the celebrated AdS/CFT correspondence \cite{Maldacena:1997re} 
(for a set of pedagogical introductions see \cite{Ramallo:2013bua,Edelstein:2009iv}). In this framework the energies of the strings give a prediction for the conformal dimensions of the dual field theory operators at strong coupling $\lambda=g_{YM}^2 N \gg 1$, a piece of information that is inaccessible from perturbative gauge theory calculations.  
A case where significant progress has been made is that of the original AdS/CFT scenario. The progress was possible due to the presence of integrability which manifests itself at both the weak and strong coupling regimes \cite{Minahan:2002ve,Bena:2003wd}. However, even in this case obtaining information about the non-planar spectrum or higher point correlation functions of gauge invariant operators is very intricate, essentially 
due to the huge mixing problem \cite{Georgiou:2008vk,Georgiou:2009tp,Georgiou:2011xj}. Recently, some progress have been also made  in this front by the use of  techniques based on Feynman diagrams \cite{Georgiou:2012zj}, integrability \cite{Basso:2015zoa} and the AdS/CFT correspondence \cite{Zarembo:2010rr,Costa:2010rz,Kazama:2013qsa,Georgiou:2010an,Georgiou:2011qk,Bajnok:2016xxu}.\footnote{Furthermore, in 
\cite{Georgiou:2016kge} the entanglement entropy of the $\mathcal {N}=4$ SYM spin chain was studied using coordinate and algebraic Bethe ansatz techniques.}

Nevertheless, there is a particularly useful and interesting double-scaling limit, called the BMN limit \cite{Berenstein:2002jq}, in which one focuses on operators with large $R$-charge. In this limit of the AdS/CFT correspondence one can solve explicitly for the string spectrum and also extract information about  structure constants involving non-protected operators.
These non-BPS operators are dual to string states propagating on the pp-wave limit of the $AdS_5 \times S^5$ background which can be obtained by focusing on the geometry around a null geodesic.
The cubic string Hamiltonian of the pp-wave background of the $AdS_5 \times S^5$ geometry has attracted a lot of attention and different proposals concerning its form  had been put forward  in \cite{Spradlin:2002ar,Pankiewicz:2002tg,DiVecchia:2003yp}.
Since the cubic Hamiltonian was not obtained from first principles the issue was how to correctly relate the string amplitudes obtained from  the pp-wave cubic Hamiltonian to the structure constants of the $\mathcal {N}=4$ SYM. This issue was resolved in \cite{Dobashi:2004nm,Lee:2004cq} by combining a number of results available from both  the string and the field theory sides \cite{Georgiou:2004ty,Georgiou:2003kt,Georgiou:2003aa,Chu:2002pd}.

More recently,  the pp-wave geometry resulting from the Penrose limit taken around a certain null geodesic
of the non-supersymmetric Schr\"{o}dinger background \cite{Alishahiha:2003ru}  was obtained in \cite{Georgiou:2019lqh}. This pp-wave background admits 16 Killing spinors and the spectrum of its bosonic excitations was derived. Compelling agreement with the dispersion relation of the giant magnons in the  original Schr\"{o}dinger spacetime obtained previously in \cite{Georgiou:2017pvi} was found. This fact led  to a conjecture of the exact, in the t'Hooft coupling, dispersion relation
of the magnons in the original  Schr\"{o}dinger background. 
 
 The aim of this work is to continue the study of the gauge/gravity correspondence relating the non-supersymmetric Schr\"{o}dinger spacetime to its dual null dipole-deformed ${\cal N}=4$ SYM \cite{Maldacena:2008wh,Herzog:2008wg,Adams:2008wt}.
 Due to the radical nature of the deformation\footnote{The generating T-s-T generating transformation introduces a mild non-locality along one of the light-cone directions of the dual gauge theory. At the same time, the dual field theory does not have a reference state and thus one can not use conventional integrability techniques (Bethe ansatze) in order to find the anomalous dimensions of the field theory operators \cite{Guica:2017jmq}.} very few observables have been calculated in the deformed version of the correspondence compared to the original AdS/CFT scenario. 
 
 Two, three and $n$-point correlation functions of scalar operators were firstly calculated in 
 \cite{Fuertes:2009ex, Volovich:2009yh} by making use of the gravity side of the correspondence. We should stress that the aforementioned operators correspond to point-like strings propagating on the $Sch_5 \times S^5$ spacetime. Extended dyonic giant magnon and spike solutions and their dispersion relations were found for the first time in
\cite{Georgiou:2017pvi}.\footnote{Giant-magnon like solutions with a different dispersion relation were studied in
\cite{Ahn:2017bio}.} Their existence is in complete agreement with the fact that the deformation does not destroy the integrability of the parent ${\cal N}=4$ SYM. 
On the field theory side, it is only the one-loop spectrum of operators belonging in a $SL(2)$ closed sub-sector that has been studied \cite{Guica:2017jmq} and agreement was found between the one-loop anomalous dimensions of certain long operators and the string theory prediction (see also \cite{Ouyang:2017yko}).

Subsequently, three-point correlation functions involving two {\it heavy} operators and a {\it light} one were calculated in \cite{Georgiou:2018zkt} using holography. The  {\it light} operator was chosen to
be one of the modes of the dilaton while the {\it heavy} states were chosen to be 
generalizations of the giant magnon or spike solutions that were constructed in \cite{Georgiou:2017pvi}. 
These results are the first in the literature where three-point correlation functions involving 
 {\it heavy} states described by extended string solutions were computed.
The aforementioned results provide the leading term of the correlators in the large $\lambda$ expansion.
They are in complete agreement with the form of the correlator dictated by the non-relativistic conformal invariance which is a symmetry of the dual theories.
Finally, certain pulsating strings solutions in the Schr\"{o}dinger background were found in \cite{Dimov:2019koi}.

Within the gauge/gravity duality framework, stable configurations of D3-branes play an important role.
Drawing inspiration from the seminal work of Myers \cite{Myers:1999ps}, in \cite{McGreevy:2000cw} 
an expanded brane configuration with the same quantum numbers as the point particle was realized. 
This is the giant graviton and it is described by a D3-brane wrapping an $S^3$ inside the $S^5$ part of the 
$AdS_5 \times S^5$ geometry.
The graviton has a nonzero angular momentum along the equator of the internal space and its stability 
against shrinking is supported by the RR repulsion. One of the highlights of the giant graviton construction 
is the explicit realization of the stringy exclusion principle. Since the giant graviton radius cannot exceed that 
of the spacetime, an upper bound for the brane momentum is implemented. In \cite{Grisaru:2000zn, Hashimoto:2000zp}
another configuration with similar properties was found, where the $S^3$ was located inside the 
$AdS_5$ part of the geometry. This is known as the dual giant graviton. 

Introducing a deformation to the geometry leads to a variety of behaviors as far as the energy difference 
between the giant and the point graviton is concerned. 
There are cases where the energy of the giant graviton is higher than 
the energy of the point graviton (see e.g. \cite{Prokushkin:2004pv, deMelloKoch:2005jg}), 
and cases where the energies  do not depend on the deformation parameter and as a result the initial degeneracy between the 
giant and the point graviton is not lifted (see e.g. \cite{Pirrone:2006iq, Imeroni:2006rb}).
To the best of our knowledge, there is a single case where the energy of the giant is lower than the energy of the point graviton (see  \cite{Huang:2007th}). We will see that our solution presents a similar behavior.

A natural follow-up question to investigate is whether the giants that were found in the original 
$AdS_5 \times S^5$ background survive the Penrose limit. 
In a pp-wave background we are trying to figure out the geometry that a particle sees/feels along the geodesic, 
after a boost to an infinite momentum frame. In the $AdS_5 \times S^5$ background that question 
was addressed affirmatively   
in \cite{Takayanagi:2002nv, Skenderis:2002vf}, while in the case of the marginally deformed ${\cal N} =4$ SYM 
of \cite{Lunin:2005jy} in \cite{Hamilton:2006ri, Avramis:2007wb, Shin:2015uwa}.
The general conclusion from the analysis of the energy of the giant graviton in the pp-wave limit of the marginally deformed 
backgrounds was that (depending on the geodesic that was under consideration) the presence of the deformation might or 
might not lift the degeneracy of the giant and the point graviton. In the cases that the degeneracy is lifted, it is
for the benefit of the point graviton. The giant graviton of \cite{Avramis:2007wb} is energetically disfavored and 
moreover, above a critical value of the deformation parameter it completely disappears from the spectrum.  

The aim of this work is to study the effect of the dipole deformation on the energy of the giant graviton and the fate 
of the degeneracy between giant and point gravitons.  For that reason we consider the pp-wave limit of the 
Schr\"{o}dinger spacetime that was constructed in \cite{Georgiou:2019lqh} and we investigate the effect of the  
deformation on the giant graviton energy that this geometry supports. 
In section \ref{section-Giant-Graviton} we revisit the pp-wave geometry, present the ansatz for the D3-brane
and calculate the energy of the point and giant gravitons. The plot of the Light-Cone Hamiltonian as a function 
of the size of the graviton reveals a rich structure when the deformation is turned on. 
In section \ref{section-fluctuations} we examine the perturbative stability of the giant graviton solution 
around the equilibrium configuration of section \ref{section-Giant-Graviton}. Enhancing the analysis of the 
scalar spectra of frequencies with the gauge field fluctuations reveals a non-trivial coupling between scalar and vector
modes, which is driven by the deformation. It is, thus, essential to include both the scalar and gauge fluctuations in the stability analysis. We conclude the paper, in section \ref{section-conclusions}. 
In appendixes \ref{second-order-terms} and \ref{quartic_coeffs} we present details of the computation of the main
text while appendix \ref{SUSY} is devoted to supersymmetry. 
We calculate the 16 killing spinors that the pp-wave background admits and use this information to 
perform a Kappa-symmetry analysis of the D3-brane probe. The analysis that is detailed in appendix 
\ref{kappa} concludes that the probe completely breaks the supersymmetry of the pp-wave background.


\section{Giant Gravitons on the pp-wave limit of the Schr\"{o}dinger geometry}
\label{section-Giant-Graviton}

We begin this section by first reviewing the pp-wave solution of the Schr\"{o}dinger geometry that was presented 
in \cite{Georgiou:2019lqh}. To obtain this solution, the Penrose limit around the null geodesic of \cite{Guica:2017jmq} 
was taken. The pp-wave background consists of the metric
\begin{eqnarray} \label{pp-metric}
 ds^2 &=& 2 \, d u \, d v
 - \Bigg[\omega^2 \left(y_3^2+y_4^2 + \rho^2 \right) + \mu^2 m^2 \left( y_3^2+y_4^2 + 4 y_1^2\right)\Bigg] du^2
 + \, \sum_{i=1}^4 (dy_i)^2
\nonumber \\[5pt]
&&
 +2 \, 
\omega \Big[y_1 \, dy_2 -y_2 \, dy_1  \Big]du + d\rho^2 + \rho^2 \Big[ d\psi^2 +\sin^2 \psi \, d\phi_2^2 +
\cos^2 \psi \, d\phi_3^2 \Big] 
\end{eqnarray}
the B-field 
\begin{equation}
B_2 \, = \,  \mu \, m \, du \wedge \Bigg[ -y_1 \wedge dy_2 + y_2 \wedge dy_1 +   
\rho^2 \Big[\sin^2 \psi \, d\phi_2 + \cos^2 \psi \, d\phi_3 \Big] \Bigg]
\end{equation}
and the RR 5-form that can be obtained as the derivative of the following potential
\begin{equation}
A_4 \, = \, \omega \Bigg[4 \, y_1 \, dy_2 \wedge dy_3 \wedge dy_4 + \rho^4 \, \cos \psi \, \sin \psi \, 
d\psi \wedge d\phi_2 \wedge d\phi_3 \Bigg] \wedge du \, . 
\end{equation}
$\mu$ and $m$ denote the deformation parameter and the eigenfunction of the mass operator of the parent Schr\"{o}dinger 
background respectively while $\omega$ is the angular velocity of the particle traveling along the null 
geodesic around which we take the Penrose limit (for more details see \cite{Georgiou:2019lqh}).

In order to describe Giant Graviton solutions, we need to consider the action of a probe D3-brane in the 
Schr\"{o}dinger pp-wave background. This is given as a sum of the Dirac-Born-Infeld (DBI) term and the Wess-Zumino (WZ) term 
\begin{equation} \label{D3-action}
S_{{\rm D}3} = - T_3 \int d^4 \xi  \, e^{-\Phi} \sqrt{\big. - \det {\cal P}\Big[g+B+2 \pi \alpha' F \Big] } +
 T_3 \int \sum_q {\cal P}\Big[ A_q \wedge e^{B+2 \pi \alpha' F} \Big]
\end{equation}
where $T_3$ is the tension of the D3-brane and ${\cal P}$ denotes the pullback of the different spacetime fields 
on the worldvolume of the brane.

We want to describe a D3-brane wrapping a three-sphere inside the undeformed part of the spacetime, in such a way that 
there is an explicit dependence of the D3-brane on-shell action from the deformation parameter $\mu$. 
For that, we consider that the brane extends along the following directions
\begin{equation}
\tau = u \, ,\qquad 
\xi_1 = \psi \, ,\qquad 
\xi_2 = \phi_2  \quad \& \quad 
\xi_3 = \phi_3
\end{equation}
and consider the following ansatz for the rest of the coordinates
\begin{equation} \label{D3-ansatz}
v = - \nu \, u \, ,\qquad 
y_1= y_3 = y_4 =0 \quad y_2 = y_0 = {\rm const}  \quad \& \quad
\rho = \rho_0 = {\rm const} \, . 
\end{equation}
Since this a not a priori consistent way of embedding the brane inside the ten-dimensional geometry, 
we need to check whether the ansatz \eqref{D3-ansatz} satisfies the equations of motion. Implementing this 
consistency check we realize that it is true only when the parameter $\nu$ acquires the following two values
\begin{equation} \label{nu-constraint}
\nu = - \, \frac{2}{9}\, \rho_0^2 \, \omega^2 \, \Big(1 \mp \Delta \Big) \, \Big(2 \pm \Delta \Big)
\quad {\rm where} \quad
\Delta \equiv \sqrt{1 \, + \, \frac{3\, \mu^2 \, m^2}{\omega^2} } \, . 
\end{equation}
These constraints for $\nu$ are coming from the equation of motion for $\rho$.
Notice that while the equations of motion for $y_1$, $y_3$ and $y_4$ are imposing the value zero to those 
coordinates, the equation of motion for $y_2$ is satisfied for any constant value of $y_2$. This in turn allows us to 
interpret the value of the constant $y_0$ as a modulus for our solution. 

The on-shell brane action, after integrating the spatial coordinates of the world volume becomes
\begin{equation} \label{on_shell_action}
S_{{\rm D}3} = \int du \, L_{{\rm D}3} 
\end{equation}
with $L_{{\rm D}3}$ being the expression for the Lagrangian
\begin{equation} \label{Lagrangian}
L_{{\rm D}3} = -\, M \, 
\Bigg[ \rho_0^3 \, \sqrt{2 \, \nu + \rho_0^2 \, \left(\omega^2 - \mu^2 \, m^2 \right)\big.} - \rho_0^4 \, \omega \Bigg]
\quad {\rm where} \quad 
M = 2 \, \pi^2 \, T_3 \, . 
\end{equation}
The conjugate momentum to $\nu$ is
\begin{equation} \label{conj_momentum}
P = - {\partial L_{{\rm D}3} \over \partial \nu}  
\quad \Rightarrow \quad 
\nu = \,- \, \frac{\rho_0^2 \, \omega^2}{2} \Bigg[1 - \frac{\mu^2 \, m^2}{\omega^2} - 
\frac{M^2 \, \rho_0^4}{\omega^2 \, P^2}  \Bigg] \, . 
\end{equation}
Calculating the light-cone Hamiltonian and substituting the expression for $\nu$  from \eqref{conj_momentum}
we arrive to an expression for the energy $E$ that is a function solely of $\rho_0$
\begin{equation} \label{LC-Hamiltonian}
E = \nu \, {\partial L_{{\rm D}3} \over \partial \nu} -L_{{\rm D}3} 
\quad \Rightarrow \quad 
E = {M^2 \over 2 \, P} \, \rho_0^6  - \omega \, M \, \rho_0^4 + \frac{P\, \omega^2 }{2}\, \left(1 - 
\frac{\mu^2 \, m^2}{\omega^2}\right)\, \rho_0^2 \, . 
\end{equation}
The local extrema of $E$ appear for the following values of the radii $\rho_0$
\begin{equation} \label{rho-local-extrema}
\rho_0 = 0  \quad \& \quad
\rho_0 = \rho_{0\pm} = \sqrt{\frac{P\, \omega}{3 \, M}\, \Big(2 \pm \Delta \Big)}
\end{equation}
while the corresponding light-cone energies become
\begin{equation}\label{dispersion-GM}
E_0 = 0 \quad \& \quad
E_{\pm} = {P^2 \, \omega^3  \over 27 \, M} \, \Big( 2 \pm \Delta \Big)^2 \,
\Big( 1 \mp \Delta \Big) \, . 
\end{equation}
Finally we substitute the values of $\rho_0$ at the two local extrema as calculated in \eqref{rho-local-extrema}
to the second of the expressions in \eqref{conj_momentum}. In that way we obtain two constraints for $\nu$
at the two extrema, which are identical to the values of $\nu$ that were calculated in \eqref{nu-constraint}.

\begin{figure}[ht] 
   \centering
   \includegraphics[width=12cm]{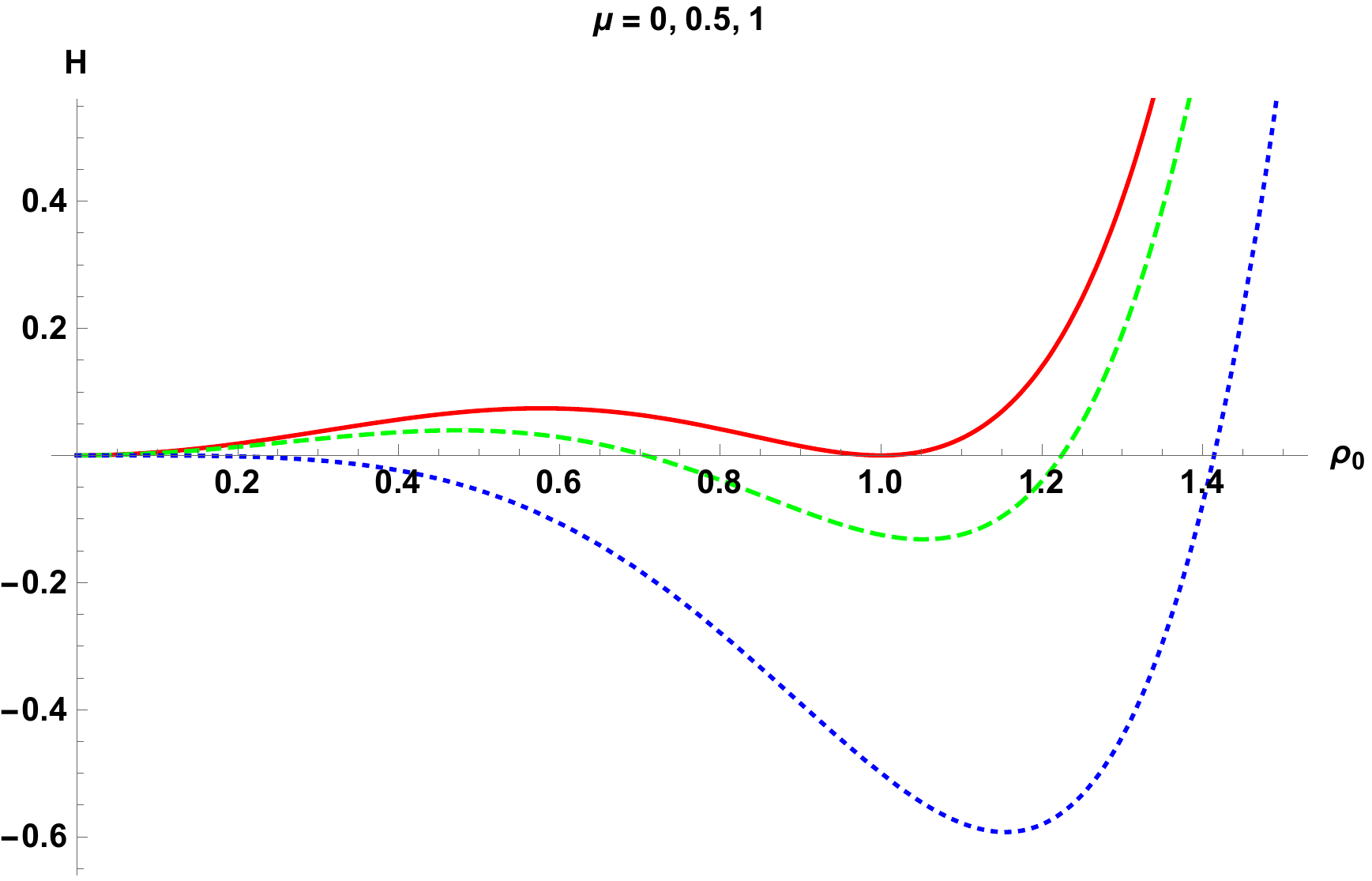}
     \caption{Light-cone Hamiltonian as a function of the graviton's size $\rho_0$. The plots are in units where $M=P=\omega=m=1$.
     The three curves correspond to three different values of the deformation parameter:  $\mu=0$ (red solid),  
     $\mu=0.5$ (green dashed) and $\mu=1$ (blue dotted). Notice that the plots are for a specific (arbitrary) 
     value of the modulus $y_0$. Changing the value of the modulus, and since the Light-cone Hamiltonian 
     does not depend on $y_0$, produces a series of degenerated vacua distributed one next to the other 
     along the extra (moduli) axis. 
     For values of the deformation that  $\mu \ge 1$ the potential has the shape of a Mexican hat but with the periodic direction decompactified and this signifies 
     the spontaneous breaking of the conformal invariance.} 
     \label{LC-Hamiltonian-plot} 
\end{figure}

In figure \ref{LC-Hamiltonian-plot} we plot the light-cone Hamiltonian for the D3-brane configuration (from  
equation \eqref{LC-Hamiltonian}) as a function of $\rho_0$. From this plot a very interesting behavior arises.

\begin{itemize}

\item $0 \le \mu < \frac{\omega}{m}$: The radii $\rho_0=0$ and  $\rho_0=\rho_{0+}$ correspond to the two local minima
(the point graviton and the giant graviton respectively), while the radius $\rho_0=\rho_{0-}$ corresponds to the 
intermediate local maximum. When the deformation is zero (red solid curve in figure \ref{LC-Hamiltonian-plot})
the giant and the point gravitons are degenerate in energy, i.e. $E_0=E_{+}=0$. Increasing the value of the 
deformation parameter $\mu$, the energy of the giant becomes less that the energy of the point graviton 
($E_{+}<E_0=0$) and the previous degeneracy is lifted. Moreover, contrary to other cases in the literature 
(see e.g. \cite{Avramis:2007wb}), the giant becomes now energetically favored with respect to the point graviton 
(green dashed curve in figure \ref{LC-Hamiltonian-plot}).

\item $\mu \ge \frac{\omega}{m}$: At the value of the deformation parameter $\mu = \frac{\omega}{m}$ 
the intermediate local maximum 
disappears and the only extrema are the minimum at $\rho_0=\rho_{0+}$ and the point graviton at $\rho_0=0$, which 
now becomes a maximum of the potential (blue dotted curve in figure \ref{LC-Hamiltonian-plot}). For values of $\mu$ 
greater or equal to this critical value the D3-brane feels a potential which has the shape of a slice of the Mexican hat
and this is suggestive of a spontaneous breaking of conformal symmetry. In figure \ref{LC-Hamiltonian-plot} we have plotted 
a projection of our potential along  a constant value of the modulus $y_0$, but combining all these projections 
(which will have the same shape for the potential since the Light-cone Hamiltonian does not depend on $y_0$) 
the complete shape of the potential will arise.  We will comment more on this feature in the
following. Furthermore, the point graviton is not part of the spectrum.

\end{itemize}

The presence of a Mexican hat-like potential means that if we insert a brane in the background and tune the 
deformation parameter to satisfy the inequality  $\mu \ge \frac{\omega}{m}$,
the brane does not have the opportunity to {\it sit} in the origin and has to choose the position $\rho_0=\rho_{0+}$. 
This in turn induces a scale in the geometry of the system (Schr\"{o}dinger background plus probe D3-brane) and suggests 
that conformal invariance is spontaneously broken.\footnote{This spontaneous breaking of the conformal invariance can be , for example, be realized holographically if one calculates the  one-point function of a BPS operator in a fashion similar to \cite{Kristjansen:2012tn}. This will be non-zero and will depend on the size of the giant graviton.} Notice that when the parameter is less than the critical value
the brane can always choose the position of the point graviton, which is a metastable state and requires an exponentially large amount of time  to reach through tunnelling the giant graviton state.

The deformation of the parent AdS space to produce the Schr\"{o}dinger background breaks a lot of 
the initial symmetry. Because of that, it is almost impossible to isolate a three-sphere inside the Schr\"{o}dinger
part of the geometry, wrap there a D3-brane and construct the dual Giant Graviton solution. 


\section{Perturbative stability of the Giant Graviton solution}
\label{section-fluctuations}

The analysis of the previous section revealed that the giant graviton is energetically favored with respect to 
the point graviton. The next important step is to examine the perturbative stability of the giant graviton solution 
around the equilibrium configuration, as first studied in \cite{Das:2000st}. 
The D3-brane vibration can be described by expanding around the previous 
ansatz  \eqref{D3-ansatz} as follows\footnote{We restrict the fluctuation analysis to the case of zero value for the modulus 
$y_0$, namely $y_0 = 0$. The shapes of the vacua for the different values of the modulus are identical, so one can argue by continuity that 
if the spectrum is free of instabilities in the zero modulus case it will remain so even when 
the modulus is different from zero.}
\begin{eqnarray} \label{fluctuation_ansatz_1}
& v \, = \,  - \, \nu \, u \, + \, \delta v \big(u,\psi,\phi_2,\phi_3 \big) \, , \quad  
\rho \, = \, \rho_0 \, + \, \delta \rho \big(u,\psi,\phi_2,\phi_3 \big) &
\nonumber \\[5pt] 
& y_i \, = \, \delta y_i \big(u,\psi,\phi_2,\phi_3 \big) \quad {\rm for} \quad i=1,2,3,4&
\end{eqnarray}
together with an expansion for the worldvolume gauge fields
\begin{equation} \label{fluctuation_ansatz_2}
A_i \, =  \, \delta A_i \big(u,\psi,\phi_2,\phi_3 \big) \quad {\rm for} \quad 
i=u, \psi, \phi_2 ,\phi_3 \, . 
\end{equation}
An important comment is in order. 
The choice of \eqref{fluctuation_ansatz_1} and \eqref{fluctuation_ansatz_2} as the perturbative ansatz for the 
giant graviton
emphasizes that we study the full bosonic spectrum; in other words we turn on both the scalar and the 
gauge field fluctuations on the worldvolume of the D3-brane. Similarly to the giant graviton fluctuations in the 
marginally deformed ${\cal N} =4$ SYM (see \cite{Pirrone:2006iq} and \cite{Pirrone:2008av}) the analysis of this
section will reveal that the Schr\"{o}dinger 
deformation of the background induces a non-trivial coupling between the scalar and the vector modes. 
This coupling in turn indicates that in order to correctly determine the dependence of the frequencies on the deformation
parameter the fluctuations of the gauge fields have to be taken account and it is not consistent to set them to zero.
 
Since we are interested for perturbations around the giant graviton solution, in the rest of the analysis of 
this section we fix $\nu$ to the following value 
\begin{equation} \label{nu_GG}
\nu = - \, \frac{2}{9}\, \rho_{0}^2 \, \omega^2 \, \Big(1 - \Delta \Big) \, \Big(2 + \Delta \Big) \, . 
\end{equation}

Subsequently, after computing the pullbacks of all the different fields and inserting them in DBI and WZ terms of 
the action \eqref{D3-action}, we expand up to second order in the fluctuations
\begin{equation}
S_{{\rm D}3} \, = \, S_0 \, + \, S_1 \, + \, S_2 + \ldots 
\end{equation}

The zeroth term in the expansion above is the classical on-shell action given in equations \eqref{on_shell_action} and 
\eqref{Lagrangian}, after substituting the value for $\nu$ from \eqref{nu_GG}
\begin{equation}
S_0 \, = \, \frac{1}{3} \, M \, \rho_0^4 \, \omega \, \Big(1 - \Delta \Big) < 0 \, . 
\end{equation}
For non-zero deformation the value of $S_0$ is always negative, highlighting the fact that the giant graviton 
is energetically favored with respect to the point graviton (for which the value of the on-shell action is zero).

The first order term in the expansion reads
\begin{equation}
S_1 \, = \,  \frac{3 \, M\, \rho_0^2}{2 \, \pi^2 \, \omega \, \big(2+ \Delta \big)}  \, 
\int d u \, d \, \Omega_3 \, \Bigg[\partial_u \delta v + \frac{\omega}{\sqrt{3}} \, \sqrt{\Delta^2-1}\, 
\Big(F_{u \phi_2}+F_{u \phi_3}\Big)\Bigg] 
\end{equation}
and it is a total derivative that vanishes upon integration by parts. 

The second order term in the expansion reads
\begin{equation} \label{action-second-order}
S_2 \, = \, \frac{M}{2\, \pi^2} \, \frac{1}{2+\Delta}\,\int d u \, d \, \Omega_3 \, \Big( {\cal S}_2^{I} +
{\cal S}_2^{II} + {\cal S}_2^{III} + {\cal S}_2^{IV} + {\cal S}_2^{V} \Big)
\end{equation}
where all the different terms comprising \eqref{action-second-order} are listed in the appendix \ref{second-order-terms}.

To further proceed with the analysis, we need to expand all the fluctuations in the basis that is spanned by combinations 
of the scalar harmonics of $S^3$. Those harmonics are eigenfunctions of the operators $\triangle_ {{\rm S}^3}$ and 
$\partial_{\phi_{2, 3}}$ simultaneously with
\begin{equation}
\triangle_ {{\rm S}^3} \Psi_{{\rm S}^3} \, = \, - \, l  \left(l +2\right) \Psi_{{\rm S}^3} 
\end{equation}
or explicitly
\begin{equation}
\frac{1}{\sin 2 \, \psi} \, \partial_{\psi} \Big( \sin 2 \, \psi \, \partial_{\psi} \Psi_{{\rm S}^3}  \Big) + \Bigg[  l \, \left(l +2\right) 
+ \frac{\partial^2_{\phi_2} }{\cos^2 \psi} \, + \, \frac{\partial^2_{\phi_3} }{\sin^2 \psi} \, \Bigg]\Psi_{{\rm S}^3} \, = \,0
\end{equation}
and
\begin{equation}
\partial_{\phi_{2, 3}}  \Psi_{{\rm S}^3} \, = \, i \, n_{2,3} \, \Psi_{{\rm S}^3}
\end{equation}
where the three quantum numbers $l, n_2$ and $n_3$ have to satisfy the following relation\footnote{An 
explicit expression of the scalar harmonics in terms of the Jacobi polynomials appears in \cite{Hernandez:2005xd, Hernandez:2005zx}.}
\begin{equation} \label{quantum-number-relation}
l \, = \, 2 \, k \, + \, |n_2| \, + \, |n_3| \quad {\rm with} \quad l=0,1, \dots \quad \& \quad k=0,1, \dots 
\end{equation}


\subsection{Scalar coupled \& decoupled modes}

We start the analysis of the fluctuations from the equation for $\delta y_3$ (or equivalently $\delta y_4$)
that is decoupled from the rest, so we consider the following ansatz
\begin{equation} \label{fluct_ansatz_y3}
\delta y_{3,4} = e^{-{i}  \, \Omega \, u} \, \Psi_{{\rm S}^3} \, .
\end{equation}
Substituting \eqref{fluct_ansatz_y3} in the equation of motion for $\delta y_3$ we arrive to an algebraic equation for the 
frequency $\Omega$ that can be solved analytically
\begin{equation}
\frac{\Omega^2}{\omega^2} \, = \, \frac{1}{9} \, \big(2+ \Delta \big)^2 
\Bigg[ l  \left(l +2\right) + \frac{3}{\big(2+ \Delta \big)^2} \, \Big[2+ \Delta^2 + \left(\Delta^2-1\right) 
\left(n_2+n_3\right)^2 \Big]\Bigg] >0 \, .
\end{equation}
A positive definite frequency implies stability against small perturbations in the direction $y_3$ 
(and equivalently $y_4$). 

The fluctuations along the directions $y_1$ and $y_2$ are coupled, so we consider the following ansatz
\begin{equation} \label{fluct_ansatz_y1_y2}
\delta y_1 = \alpha_1 \, e^{-{i}  \, \Omega \, u} \, \Psi_{{\rm S}^3} \quad \& \quad  
\delta y_2 = \alpha_2 \, e^{-{i} \, \Omega \, u} \Psi_{{\rm S}^3}
\end{equation}
where $\alpha_1$ and $\alpha_2$ are constants. Substituting  \eqref{fluct_ansatz_y1_y2} in the 
equations for $y_1$ and $y_2$,  an algebraic coupled
system for $\alpha_1$ and $\alpha_2$ arrises. Imposing the determinant to vanish, we arrive to 
following depressed quartic equation for the frequencies
\begin{equation} \label{quartic}
\Omega^4 \, + \, \gamma_2 \, \Omega^2 \, + \, \gamma_1 \, \Omega \, + \gamma_0 \, = \, 0 
\end{equation}
where the value for the constants $\gamma_2$, $\gamma_1$ and $\gamma_0$ are listed in appendix 
\ref{quartic_coeffs}. Since all the roots of the quartic equation \eqref{quartic} are real numbers, 
this suggests stability of the spectrum along the directions $y_1$ and $y_2$.


\subsection{Scalar \& gauge coupled modes}

The equations of motion for the two scalars $\delta v$, $\delta \rho$ and the four gauge fields $\delta A_u$, 
$\delta A_{\psi}$, $\delta A_{\phi_2}$ and $\delta A_{\phi_3}$ are coupled. 
There are two types of proposed ans\"{a}tze, one corresponding to the Type II mesons of \cite{Kruczenski:2003be} 
(see also \cite{Itsios:2016ooc}) and a second one corresponding to the Type III mesons of the same paper.
For the Type II fluctuations the ansatz is
\begin{equation} \label{ansatz_TypeII}
\delta v = \alpha_v \, e^{-i  \, \Omega \, u} \, \Psi_{{\rm S}^3} \, , \quad  
\delta \rho = \alpha_{\rho} \, e^{-{ i} \, \Omega \, u} \Psi_{{\rm S}^3} \, , \quad 
\delta A_u = \alpha_u \,  \Psi_{{\rm S}^3} \quad \&  \quad 
\delta A_{i} =0\, ,
\end{equation}
where in the above and the following equation 
$i= \psi, \phi_2 ,\phi_3 \, $. 
For the Type III fluctuations the ansatz is
\begin{equation} \label{ansatz_TypeIII}
\delta v = \alpha_v \, e^{-i  \, \Omega \, u} \, \Psi_{{\rm S}^3} \, , \quad  
\delta \rho = \alpha_{\rho} \, e^{-{ i} \, \Omega \, u} \Psi_{{\rm S}^3} \, , \quad 
\delta A_u = 0 \, , \quad  
\delta A_{i} = \alpha_{III} \, e^{-i  \, \Omega \, u} \, \nabla_i \Psi_{{\rm S}^3}
\end{equation}
with $ \nabla_i \Psi_{{\rm S}^3} $ being one class of the vector spherical harmonics.
Substituting the Type II ansatz  of \eqref{ansatz_TypeII} in the coupled system of equations of motion, 
we obtain a constraint for the constant $\alpha_u$ 
\begin{equation} \label{constraint-II}
\alpha_u \, = \, -\, \frac{\omega}{\sqrt{3} \, \Omega}\, \sqrt{\Delta^2 - 1}  \, \big(n_2 + n_3 \big) \, \alpha_v
\end{equation}
and an algebraic coupled system for the $\alpha_v$ and $\alpha_{\rho}$. Imposing the vanishing of the 
determinant we arrive to a quartic equation for $\Omega$ with solutions 
\begin{equation}
\frac{\Omega_{+}^2}{\omega^2} \, = \ \frac{1}{3} \, \left(\Delta^2-1\right) \left(n_2+n_3\right)^2 \, + \,
\frac{1}{9} \, \left(l +2\right) \, \big(2+ \Delta \big) \Big[6 \, \Delta + l \, \big(2+ \Delta \big) \Big]
\end{equation}
and 
\begin{equation}
\frac{\Omega_{-}^2}{\omega^2} \, = \ \frac{1}{3} \, \left(\Delta^2-1\right) \left(n_2+n_3\right)^2 \, + \,
\frac{1}{9} \, l  \, \big(2+ \Delta \big) \Big[4 \, \big(1- \Delta \big) + l \, \big(2+ \Delta \big) \Big] \, .
\end{equation}
Following the same strategy for the Type III ansatz  of \eqref{ansatz_TypeIII}, we obtain a
constraint  for the constant $\alpha_{III}$
\begin{equation} \label{constraint-III}
\alpha_{III} \, = \, \frac{i}{\sqrt{3} \, \Omega^2} \, \sqrt{\Delta^2 - 1} \, \big(n_2 + n_3 \big) \, \alpha_v
\end{equation}
while arriving to the same frequencies as to the Type II fluctuations.

The frequencies $\Omega_{+}^2$ are positive definite for any value of the deformation and the only potential source 
of instability, along these coupled fluctuations, might come if $\Omega_{-}^2$ becomes negative.  
This would be possible if the following inequality is satisfied for $\Delta \ge 1$ 
\begin{equation} \label{inequality}
\Delta^2 \, \left[\big(n_2 + n_3 \big) ^2 + \, \frac{l}{3} \, \left(l-4\right)\right] + \frac{4}{3}\,  \Delta \, l \,  \left(l-1\right)
+ \frac{4}{3}\, l \,\left(l+2\right) -  \big(n_2 + n_3 \big) ^2 < 0 \, .
\end{equation}
Examining \eqref{inequality} for different values of $n_2, n_3$ and $l$ and taking into account the 
relation \eqref{quantum-number-relation}, we conclude that 
the only case where \eqref{inequality}  is satisfied is for $n_2=n_3=0$, $l=2$ and $\Delta>4$. 
For values of the deformation in the interval  $1 \le \Delta \le 4$ inequality \eqref{inequality} is never satisfied for 
choices of the quantum numbers  $n_2, n_3$ and $l$ that are allowed by \eqref{quantum-number-relation}.

It is interesting to notice that the only potential source of instability for the coupled modes of this subsection 
is coming when we constrain the ansatze 
\eqref{ansatz_TypeII} and \eqref{ansatz_TypeIII} to $u$ and $\psi$ dependence (avoiding the dependence on the 
angles $\phi_2$ and $\phi_3$). The effect of such a choice (see \eqref{constraint-II} and \eqref{constraint-III})
is that we {\it effectively} decouple the gauge field sector. Therefore the conclusion from the above analysis 
is that in order to have a stable giant graviton (for all the values of the deformation parameter) we need a coupling
between the scalar and the gauge field sector.  Decoupling may provide a potential source of instability.


\section{Conclusions and future directions}
\label{section-conclusions}

In this paper we focus on the recently constructed pp-wave limit of the Schr\"{o}dinger spacetime \cite{Georgiou:2019lqh} and 
investigate to the direction of identifying a giant graviton solution. 
Our findings conclude that such a solution exists and moreover reveals an unexpected behavior, as the 
deformation parameter is varied. 
The degeneracy that is inherited to the pp-wave geometry from the parent $AdS_5 \times S^5$ background is lifted 
by the deformation, but for the benefit of the giant graviton. A lift of the degeneracy could be expected 
on general grounds from similar computations in the marginally deformed backgrounds, 
but the direction is clearly unexpected and for that reason very interesting.

The results of this work could be combined with those of \cite{Huang:2007th} 
where energetically favored giant gravitons were reported,
in a background that was dual to a non-commutative field theory. 
From one side, when the deformation does not affect the $AdS$ space (e.g. marginally deformed 
backgrounds) the giant is energetically disfavored and from the opposite side when 
the $AdS$ space is deformed (e.g. the deformation we study in this paper and the case of \cite{Huang:2007th}) the 
giant is energetically favored. This suggests that in order to have an energetically 
favored giant graviton with respect to the point graviton, we have to deform the $AdS_5 \times S^5$ background 
in such a way that the deformation involves $AdS$ components.\footnote{In our case this can be easily seen from \eqref{pp-metric} in which the deformation $\mu$ involves the coordinates $y_i, i=1,...,4$ which originate from the Schr\"{o}dinger (deformed $AdS_5$) part of the background.}
Moreover, in our case we have the additional feature that when the deformation parameter exceeds a certain critical value the shape of potential that the D3-brane feels is similar to that of a Mexican hat with the periodic direction decompactified. 
Thus, for strongly deformed backgrounds the mere presence of a D3-brane leads to a spontaneous breaking of conformal invariance.

Besides identifying the point and the giant graviton solutions we perform a detailed study of the full bosonic spectrum.
We take into account both the scalar and the gauge field fluctuations and from the analysis we conclude that it is 
inconsistent to restrict the computation only in the scalar sector, due to a coupling that is induced by the the deformation. 
Moreover, we realize that the coupling that the deformation provides is vital for the stability analysis of the spectrum around the giant graviton solution.
If we effectively decouple (by an appropriate choice of the quantum numbers) the gauge field sector, then 
a potential instability for the spectrum arises. However, there is a critical value of the deformation parameter below which the D3-brane is stable for all values of the 3-sphere's quantum numbers.

Finally we performed the Kappa-symmetry analysis of the D3-brane probe to check 
whether the giant graviton is a BPS object. With the computation that is presented in the appendix, we 
conclude that the probe completely breaks the supersymmetry of the pp-wave background. 

A number of future directions remain to be addressed. The first concerns the form of the field theory 
operators which are dual to the giant gravitons of \eqref{D3-ansatz}. 
It would be certainly interesting to determine their structure  and see if one can reproduce from this their dispersion relation \eqref{dispersion-GM}  at least for small values of $\frac{P^2}{M}$.
A second direction could be to find the instanton solutions and the corresponding transition probabilities from the point graviton to the energetically favored giant graviton as a function of the deformation parameter $\mu$, for the values of the deformation parameter that this tunnelling makes sense, that is 
for $\mu<\frac{\omega}{m}$. Finally, it would be interesting to consider spacetimes where both the the 
$AdS_5$ and  $S^5$ part of the background are deformed since in this case there will be two competing tendencies. On one hand, the  $AdS_5$ deformation will tend to lower the energy of the giant graviton with respect to the point graviton, while on the other hand the  $S^5$ deformation will tend to do the opposite.


\section*{Acknowledgments}

The work of this project has received funding from the Hellenic Foundation
for Research and Innovation (HFRI) and the General Secretariat for Research and Technology (GSRT), under grant agreement No 15425.


\appendix


\section{Second order terms in the expansion of the action}
\label{second-order-terms}

In this appendix we gather all the different terms of the second order expansion of the action 
in \eqref{action-second-order}. 

To facilitate the presentation we introduce the metric ${\cal G}_{ab}$ with the following non-vanishing entries
\begin{equation} \label{def-G}
{\cal G}_{uu} \, = \, - \left(\frac{2 + \Delta}{3}\right)^2 \, \omega^2\, , \quad 
{\cal G}_{\psi \psi} \, = \, 1\, , \quad 
{\cal G}_{\phi_2 \phi_2} \, = \,\sin^2 \psi  \quad \& \quad
{\cal G}_{\phi_3 \phi_3} \, = \, \cos^2 \psi 
\end{equation}
and the vector $q$ with the following non-vanishing components
\begin{equation}
q^{\phi_2} \, = \, q^{\phi_3} \, = \, \sqrt{3} \, \frac{\sqrt{\Delta^2 -1}}{\Delta+2} \, . 
\end{equation}

The terms appearing in the expansion of  \eqref{action-second-order} are
\begin{equation}
\frac{{\cal S}_2^{I}}{\omega \, \rho_0^2 } \, = \,- \, 4 \,  \big(1 - \Delta \big) \, \delta \rho^2 \, + \, 
2 \, \big(1 - \Delta^2 \big) \, \delta y_1^2 \, - \, \left(1 + \frac{\Delta^2}{2} \right) \Big( \delta y_3^2 +  \delta y_4^2\Big)
\end{equation}

\begin{equation}
{\cal S}_2^{II}\, = \, \frac{18 \, \Delta \, \rho_0}{\big(2+ \Delta \big) \, \omega} \, \delta \rho \, \partial_{u} \delta v \, - \, 
6 \, \rho_0^2 \, \delta y_2 \, \partial_{u} \delta y_1
\end{equation}

\begin{equation}
{\cal S}_2^{III}\, = \, - \, \frac{3}{2 \, \omega} \, {\cal G}^{\alpha \beta} \,  \partial_{\alpha} \delta v \,  \partial_{\beta} \delta v \, - \, 
\frac{1}{6} \, \big(2 + \Delta \big)^2 \, \omega \, \rho_0^2 \, {\cal G}^{\alpha \beta} \,  \partial_{\alpha} \delta {\vec r_5} \,  
\partial_{\beta} \delta {\vec r_5}
\end{equation}

\begin{equation}
\frac{{\cal S}_2^{VI}}{\frac{1 - \Delta^2}{2} \, \omega \, \rho_0^2}\, = \, \Big[ \big( \partial_{\phi_2} + \partial_{\phi_3} \big) \,  
\delta {\vec r_5}  \Big]^2 \, - \, 4 \, \delta y_2 \, \Big[ \big( \partial_{\phi_2} + \partial_{\phi_3} \big) \,  \delta y_1 \Big]
\end{equation}

\begin{equation}
{\cal S}_2^{V}\, = \, - \, \frac{\omega}{12} \, \big(2 + \Delta \big)^2 \, F_{ab} \, F^{ab} \, - \, 
\big(2 + \Delta \big) \, q^c \, F_{a c} \, {\cal C}^{a b} \, \partial_{b} \delta v \, + \, 6 \, \Delta \, \rho_0 \, q^c \, F_{u c} \, \delta \rho 
\end{equation}
where the matrix ${\cal C}^{ab}$ is defined as follows
\begin{equation}
{\cal C}^{ab} \, \equiv \, {\cal G}^{ab} \, + \, q^a \, q^b \quad {\rm with} \quad
F^{ab} \, = \, {\cal C}^{ac} \, {\cal C}^{b d} \, F_{cd} 
\end{equation}
and we have introduced the shorthand: 
$\delta \vec{r}_5 = \left(\delta \rho, \delta \vec{y_4}\right)$.


\section{Coefficients of the quartic equation}
\label{quartic_coeffs}

In this appendix we gather the values for coefficients of the quartic equation \eqref{quartic} that calculates the 
frequencies along the directions $y_1$ and $y_2$. 
The value for the constant $\gamma_2$ is
\begin{equation}
\gamma_2 \, = \, - \frac{2\, \omega^2}{3} \, \Bigg[\frac{1}{3} \left(2+ \Delta \right)^2  l  \left(l +2\right)  +
2 \left(\Delta^2+2 \right) + \left(\Delta^2-1\right) \left(n_2+n_3\right)^2 \Bigg]
\end{equation}
for the constant $\gamma_1$ is
\begin{equation}
\gamma_1 \, = \, - \frac{8}{3} \, \omega^3 \left(\Delta^2-1\right) \left(n_2+n_3\right)
\end{equation}
and for the constant $\gamma_0$ is
\begin{equation}
\frac{\gamma_0}{\frac{\omega^4}{81}} = 12 \, l  \left(l +2\right)  \left(2+ \Delta \right)^2 \left(\Delta^2-1\right)  +
\Bigg[ l  \left(l +2\right)  \left(2+ \Delta \right)^2 + 3 \,\left(\Delta^2-1\right) \left(n_2+n_3\right)^2\Bigg]^2 . 
\end{equation}
%


\section{Supersymmetry}
\label{SUSY}

In this appendix we will analyze the supersymmetry behavior of the Giant Graviton configuration we 
considered in section \ref{section-Giant-Graviton}. We will realize that these probe brane configurations break the 
16 supersymmetries of the pp-wave background.  

The number of supersymmetries that a Dp-brane preserves, is the number of independent solutions 
of the kappa symmetry condition
\begin{equation} \label{kappa-symmetry-equation}
\Gamma_{\kappa}\,\epsilon\,=\,\epsilon
\end{equation}
where $\epsilon$ is the Killing spinor of the background and for a Dp-brane (in a type IIB solution) $\Gamma_{\kappa}$  
is given by the following expression
\begin{equation} \label{kappa-symmetry-matrix}
\Gamma_{\kappa}\,=\,{1\over \sqrt{-{\rm det}\, {\cal P} \Big[g+B  +  2 \, \pi \, \alpha \, F \Big]}}\,\,
\sum_{n=0}^{\infty}\,{1\over 2^n\,n!}\,\,
\gamma^{\mu_1\nu_1\cdots\mu_n\nu_n}\,\, {\cal F}_{\mu_1\nu_1}\,\cdots\,
{\cal F}_{\mu_n\nu_n}\,\,J^{(n)}
\end{equation}
where ${\cal F} \, = \,{\cal P} [B \, + \, 2 \, \pi \, \alpha' \, F]$ and $J^{(n)}$ is the matrix
\begin{equation}
J^{(n)}\,=\,(-1)^n\,\sigma_3^{{p-3\over
2}+n}\,(i \, \sigma_2)\,\otimes\,\Sigma_0 
\quad \rm{with} \quad 
\Sigma_0\,=\,{1\over (p+1)!}\,\,\epsilon^{\mu_1\cdots\mu_{p+1}}\,
\gamma_{\mu_1\cdots\mu_{p+1}} \, . 
\end{equation}
Moreover, $\gamma_{\mu_1\mu_2\cdots}$ are
antisymmetrized products of the induced worldvolume Dirac matrices and the killing spinor
consists of two Majorana-Weyl spinors, which can be arranged in a two-component vector.


\subsection{Killing spinor of the pp-wave background}
\label{Killing}

The supersymmetric variations of the dilatino and the gravitino in a type IIB solution are
\begin{eqnarray} \label{SUSY-IIB}
&& \delta \lambda \, = \, \frac{1}{2} \slashed \partial \Phi \, \epsilon - \frac{1}{24} \slashed{H} \, \sigma^3 \, \epsilon
+ \frac{1}{2} e^{\Phi} \biggl[ \slashed{F}_{1} \left(i \sigma^2\right) + \frac{1}{12}\slashed{F}_3  \, \sigma^1 \biggr] 
\epsilon
\nonumber \\
&& \delta \psi_{\mu} \, = \, D_{\mu} \epsilon - \frac{1}{8} H_{\mu \nu \rho} \, \Gamma^{\nu \rho} \, \sigma^3 \, \epsilon
- \frac{e^{\Phi}}{8} \biggl[ \slashed{F}_1 \left(i \sigma^2\right) + {1\over 6} \slashed{F}_3 \sigma^1 +  
\frac{1}{2 \times  5!} \slashed{F}_{5} \left(i \sigma^2\right) \biggr]  \Gamma_{\mu} \, \epsilon 
\end{eqnarray}
where we use the notation $\slashed{F}_n \equiv  F_{i_1\dots i_n} \Gamma^{i_1 \dots i_n}$ and 
\begin{equation} \label{def-covariant}
D_{\mu} \epsilon \, = \,  \partial_{\mu} \epsilon + \frac{1}{4} \, \omega^{a b}_{\mu} 
\, \Gamma_{a b} \, \epsilon \, . 
\end{equation}
To analyze the supersymmetry transformations of \eqref{SUSY-IIB} we define the orthonormal basis
\begin{eqnarray} \label{frame}
&&
e^{-} = dv \, +  \frac{1}{2} \, {\cal H}  \, du  + \omega \left(y_1 \, dy_2 - y_2 \, dy_1 \right) \, ,
\quad
e^{i} =  dy_i  \quad  \text{for} \quad i=1\ldots 4 
\nonumber \\[10pt]
&& 
e^{+} =  du \, , 
\quad
e^{5} =  d\rho  \, , 
\quad 
e^{6} =  \rho \,  d\psi \, , 
\quad 
e^{7}  = \rho \, \sin\psi \,  d\phi_2 \, , 
\quad 
e^{8} = \rho \, \cos\psi \,  d\phi_3 \, . 
\end{eqnarray}
with
\begin{equation}
{\cal H} = - \omega^2 \left(y_3^2+y_4^2 + \rho^2 \right) - \mu^2 m^2 \left( y_3^2+y_4^2 + 4 y_1^2\right)\, . 
\end{equation}
The non-vanishing components of the spin connection are
\begin{eqnarray} \label{spin-connection}
&&
\omega_{+ 1} \, = \, \frac{1}{2} \, \partial_{y_1} {\cal H} \, e^{+} + \omega \, e^{2} \, , \quad 
\omega_{+ 2} \, = \, -\, \omega \, e^{1} \, , \quad 
\omega_{+ 3} \, = \, \frac{1}{2} \, \partial_{y_3} {\cal H} \, e^{+} \, ,
\nonumber \\[5pt]
&& 
\omega_{+ 4} \, = \, \frac{1}{2} \, \partial_{y_4} {\cal H} \, e^{+} \, , \quad
\omega_{1 2} \, = \, - \, \omega \, e^{+} \, , \quad
\omega_{+ 5} \, = \, - \, \rho \, \omega^2 \, e^{+} \, ,
\\[5pt]
&& 
\omega_{5 i} \, = \, - \, {1 \over \rho} \, e^{i} \quad  \text{for} \quad i=6,7,8   \, \quad 
\omega_{6 7} \, = \, - \, {1 \over \rho} \, \cot \psi \, e^{7}  \, \quad 
\omega_{6 8} \, = \, {1 \over \rho} \, \tan \psi \, e^{8} \, . 
\nonumber
\end{eqnarray}
Moreover, the expressions of $H_3$ and $F_5$ in frame components are
\begin{eqnarray} \label{H3F5-FrameComponents}
 H_3 &=& 2 \, \mu \, m \, e^{+} \wedge 
 \Bigg[ e^{1} \wedge e^{2} - \sin\psi \Big(e^{5} \wedge e^{7} - e^{6} \wedge e^{8} \Big)
 - \cos\psi \Big(e^{5} \wedge e^{8} + e^{6} \wedge e^{7} \Big) \Bigg]
\nonumber \\[5pt]
F_5& = & 4\, \omega \, e^{+} \wedge  \Big(  e^{1} \wedge e^{2}  \wedge e^{3} \wedge e^{4} 
+ e^{5} \wedge e^{6}  \wedge e^{7} \wedge e^{8} \Big) \, . 
\end{eqnarray}
Plugging all the ingredients in the dilatino equation and requiring that it vanishes, we arrive to the following expression
\begin{equation} \label{dilatino-Projection}
\Gamma^{+} \Big[ \Gamma^{12} \, - \, \Gamma^{58} \, - \, \Gamma^{67} \Big] \, \sigma^3 \, e^{-\frac{\psi}{2}\, 
\Gamma_{56}} \, \epsilon \, = \, 0 \, . 
\end{equation}
The equation above is satisfied only when the following projection is implemented 
\begin{equation} \label{Gp-Projection}
\Gamma^{+} \, \epsilon \, = \, 0 \, . 
\end{equation}
To determine the precise form of the Killing spinor we need to analyze the gravitino equation of \eqref{SUSY-IIB} 
for each one of the frame components, as they are defined in \eqref{frame}. It can be checked explicitly
that the spinor is independent of the variables $v$, $y_1$, $y_2$, $y_3$, $y_4$ and $\rho$ and 
depends on $\psi$, $\phi_2$ and $\phi_3$ as follows
\begin{equation} \label{spinor-1}
\epsilon \, = \, e^{\frac{\psi}{2}\, \Gamma_{56}} \,  
e^{\frac{\phi_2}{2}\, \Gamma_{67}} \, 
e^{\frac{\phi_3}{2}\, \Gamma_{58}} \, {\hat \epsilon} (u) \, . 
\end{equation}
The dependence on the variable $u$ will come from the gravitino variation along the $e^{+}$ direction.
In that way we obtain the following differential equation that ${\hat \epsilon}(u)$ should obey
\begin{equation}  \label{spinor-diff-equation}
\partial_u {\hat \epsilon} \,- \, \frac{1}{2} \, \omega \, \Gamma^{12} \, {\hat \epsilon}  \, - \, 
\frac{1}{2}\, \mu \, m \left[\Gamma^{12} \, - \, \Gamma^{58} \, -  \,  \Gamma^{67}\right] \, \sigma^3 \, {\hat \epsilon}
 \, - \,  \frac{i}{2} \, \omega \, \Big[\Pi \, + \, \Pi' \Big]  \, \sigma^2 \, {\hat \epsilon} \, = 0
\end{equation}
where we have defined 
\begin{equation} \label{Pi-definition}
\Pi \, \equiv \, \Gamma^{1234}  
\qquad \& \qquad 
\Pi' \, \equiv \, \Gamma^{5678}  \, . 
\end{equation}
Equation \eqref{spinor-diff-equation} can be solved in a similar way to the analogous equation in \cite{Georgiou:2019lqh}.


\subsection{Kappa-symmetry analysis of the D3-brane probe}
\label{kappa}

Armed with the machinery of the previous two subsections we are ready to determine the amount of supersymmetry 
that is preserved by the Giant Graviton configuration of section \ref{section-Giant-Graviton}. 
The D3-brane extends along the 
directions $u, \psi, \phi_2$ and $\phi_3$ and the $\kappa$-symmetry matrix is obtained from \eqref{kappa-symmetry-matrix}
for the value $p=3$. The induced Dirac matrices, for the embedding  \eqref{D3-ansatz}, are
\begin{eqnarray} \label{WV-gamma-down}
&& \gamma_{\tau} \, = \, \Gamma_{+} \,+ \, \left(\frac{1}{2} \, {\cal H} - \, \nu \right)\, \Gamma_{-} 
\, = \, \Gamma_{+} \, - \, \frac{\rho_0^2 \, \omega^2}{18} \, \left(1 \,+ \,  2 \, \Delta \right)^2 \, \Gamma_{-} \, , 
\nonumber \\[5pt]
&& \gamma_{\xi_1} \, = \, \rho_0 \,  \Gamma_{6} \, , \quad 
\gamma_{\xi_2} \, = \, \rho_0 \, \sin \psi \,  \Gamma_{7} \quad \& \quad  
\gamma_{\xi_3} \, = \, \rho_0 \, \cos \psi \,  \Gamma_{8} \, . 
\end{eqnarray}
Moreover 
\begin{equation} \label{B2-pb}
B_2 \, = \, \frac{1}{\sqrt{3}} \, \sqrt{\Delta^2-1} \, \rho_0^2 \, \omega \, \Big[\sin^2 \psi \, du \wedge d\phi_2 \, + \, 
\cos^2 \psi \, du \wedge d\phi_3 \Big]
\end{equation}
and 
\begin{equation} \label{sqrt-det-pb}
\sqrt{-{\rm det}\, {\cal P} \Big[g+B\Big]} \, = \, \frac{1}{3} \, \left(\Delta+2 \right)\, \rho_0^4 \, \, \omega \, \sin \psi \, \cos \psi \end{equation}
while the induced Dirac matrices with the indices up are 
\begin{eqnarray} \label{WV-gamma-up}
&& \gamma^{\tau} \, = \, \frac{1}{2} \, \Gamma_{-} \, - \, \frac{9}{\rho_0^2 \, \omega^2\, \left(1 \,+ \,  2 \, \Delta \right)^2} \, \Gamma_{+} \, , 
\nonumber \\[5pt]
&& \gamma^{\xi_1} \, = \, \frac{1}{\rho_0} \,  \Gamma_{6} \, , \quad 
\gamma^{\xi_2} \, = \, \frac{1}{\rho_0 \, \sin \psi} \,  \Gamma_{7} \quad \& \quad  
\gamma^{\xi_3} \, = \, \frac{1}{\rho_0 \, \cos \psi} \,  \Gamma_{8} \, . 
\end{eqnarray}
Substituting the values of the induced Dirac matrices (equations \eqref{WV-gamma-down} \& 
\eqref{WV-gamma-up}), $B_2$ (equation \eqref{B2-pb}) and the square root of the determinant 
(equation \eqref{sqrt-det-pb}) and using the projection \eqref{Gp-Projection}, 
the expression for the matrix $\Gamma_{\kappa}$ becomes
\begin{equation} \label{Gamma-k-3-v1}
\Gamma_{\kappa} \, = \, \frac{3}{\rho_0 \, \omega \, \left(\Delta +2 \right)} \, \Bigg[i \, \sigma_2 \, \Gamma_{+678}  - 
\sigma_1 \, \sqrt{\frac{\Delta^2-1}{3}} \, \rho_0 \, \omega \, \Gamma_{67} \, e^{-\psi\, \Gamma_{78}} \Bigg] \, .
\end{equation}
In the next step we will compute the action of $\Gamma_{\kappa}$ on the spinor of the background, that appears 
on equation \eqref{spinor-1}. By using this representation, one immediately concludes that an extra projection 
is needed, namely 
\begin{equation} \label{kappa-projection}
\Gamma_{56} \, {\hat \epsilon} \, = \, \Gamma_{78} \, {\hat \epsilon} \, \quad \Rightarrow \quad 
\Gamma_{58} \, {\hat \epsilon} \, = \, \Gamma_{67} \, {\hat \epsilon} \, . 
\end{equation}
Using \eqref{Gamma-k-3-v1}, \eqref{kappa-projection} and  \eqref{kappa-symmetry-equation} we arrive to the 
following equation that the spinor ${\hat \epsilon}$ should satisfy
\begin{equation}
{\hat \epsilon} \, = \, \frac{3}{\rho_0 \, \omega \, \left(\Delta +2 \right)} \, \Bigg[i \, \sigma_2 \, \Gamma_{+678}  - 
\sigma_1 \, \sqrt{\frac{\Delta^2-1}{3}} \, \rho_0 \, \omega \, \Gamma_{67} \Bigg] \, {\hat \epsilon}  \, .
\end{equation}
Acting from the left with $\Gamma_{+}$ (see also the supersymmetry analysis of \cite{Takayanagi:2002nv}) 
and using the fact that $\Gamma_+^2=0$ we have
\begin{equation}
\Gamma_{+} \, \Bigg[1\, + \, \frac{\sqrt{3 \big(\Delta^2-1\big)}}{\Delta +2} \, \sigma_1 \, \Gamma_{67}\Bigg] \, 
{\hat \epsilon} \, = \, 0 \, . 
\end{equation}
In order to avoid the projection $\Gamma_+ \chi = 0$, the quantity inside the brackets acting on the 
spinor has vanish.  This means that we should demand the following projection
\begin{equation}
 \Gamma_{67}  \, {\hat \epsilon}  \, = \, - \, \frac{\Delta +2}{\sqrt{3 \big(\Delta^2-1\big)}} \, \sigma_1 \, {\hat \epsilon}  \, . 
\end{equation}
However this is impossible, since acting with $\Gamma_{67}$ from the left to the above expression the result of the 
LHS is minus one while that of the RHS is not.
As a result we are forced to impose the projection $\Gamma_+ {\hat \epsilon} = 0$ and, 
in combination with \eqref{Gp-Projection},  supersymmetry in completely broken.



\end{document}